\documentclass[aps,10pt,twocolumn,superscriptaddress,preprintnumbers,amsmath,amssymb,prl]{revtex4-1}
\usepackage{times}
\usepackage[pdftex]{graphicx}
\usepackage[bookmarksopen=true,bookmarksopenlevel=1]{hyperref}
\usepackage{epstopdf}
\usepackage{amsmath}
\usepackage{amsfonts}
\usepackage{amssymb}
\usepackage{graphicx}
\usepackage{sidecap}
\sidecaptionvpos{figure}{c}
\usepackage{color}
\usepackage{dcolumn}
\usepackage{bm}
\usepackage[compact]{titlesec}
\usepackage[normalem]{ulem}
\usepackage{bookmark}
\usepackage{natbib}
\usepackage{indentfirst}
\usepackage{appendix}
\usepackage{upgreek}

\definecolor{mypink}{RGB}{219, 48, 122}
\definecolor{mygreen}{RGB}{51, 153, 102}
\definecolor{brown}{RGB}{165, 42, 42}

\newlength{\beforesection}
\newlength{\aftersection}
\newlength{\beforesubsection}
\newlength{\aftersubsection}
\titlespacing*{\section}{0pt}{\beforesection}{\aftersection}
\titlespacing*{\subsection}{0pt}{\beforesubsection}{\aftersubsection}

\def\Ef{$E_{\rm F}$}

\def\Ef{$E_{\rm F}$}

\def\invA{\AA$^{-1}$}
\def\Kbar{$\overline{\mbox{K}}$}
\def\Kbarprime{${\overline{\mbox{K}}}\!^{\,\prime}$}

\def\pntG{$\overline{\rm \Gamma}$}

\def\invA{\AA$^{-1}$}

\def\root33{$\sqrt{3}\times\sqrt{3}$ {\it R}30$^\circ$}
\def\RT3{$\sqrt{3}$}

   \makeatletter
\def\@fnsymbol#1{\ensuremath{\ifcase#1\or \dagger\or  \ast\or
   \mathsection\or \mathparagraph\or \|**\or \or \ast\ast
   \or \ast\ast \else\@ctrerr\fi}}
    \makeatother

\begin{document}

\title{Rashba-like spin textures in graphene promoted by ferromagnet-mediated electronic hybridization with a heavy metal}

\author{Beatriz Mu\~{n}iz Cano}
\altaffiliation{These authors contributed equally to this work.}
\affiliation{IMDEA Nanoscience, C/ Faraday 9, Campus de Cantoblanco, 28049 Madrid, Spain}

\author{Adri\'an Gud\'in}
\altaffiliation{These authors contributed equally to this work.}
\affiliation{IMDEA Nanoscience, C/ Faraday 9, Campus de Cantoblanco, 28049 Madrid, Spain}

\author{Jaime S\'anchez-Barriga}
\affiliation{IMDEA Nanoscience, C/ Faraday 9, Campus de Cantoblanco, 28049 Madrid, Spain}
\affiliation{Helmholtz-Zentrum Berlin f\"ur Materialien und Energie, Albert-Einstein-Str. 15, 12489 Berlin, Germany}

\author{Oliver J. Clark}
\affiliation{Helmholtz-Zentrum Berlin f\"ur Materialien und Energie, Albert-Einstein-Str. 15, 12489 Berlin, Germany}

\author{Alberto Anad\'on}
\affiliation{IMDEA Nanoscience, C/ Faraday 9, Campus de Cantoblanco, 28049 Madrid, Spain}

\author{Jose Manuel D\'iez}
\affiliation{IMDEA Nanoscience, C/ Faraday 9, Campus de Cantoblanco, 28049 Madrid, Spain}
\affiliation{Departamento de F\'isica de la Materia Condensada, Instituto Nicol\'as Cabrera and Condensed Matter \\ 
Physics Center (IFIMAC), Universidad Aut\'onoma de Madrid, Campus de Cantoblanco, 28049 Madrid, Spain}

\author{Pablo Olleros-Rodr\'iguez}
\affiliation{IMDEA Nanoscience, C/ Faraday 9, Campus de Cantoblanco, 28049 Madrid, Spain}

\author{Fernando Ajejas}
\affiliation{IMDEA Nanoscience, C/ Faraday 9, Campus de Cantoblanco, 28049 Madrid, Spain}

\author{Iciar Arnay}
\affiliation{IMDEA Nanoscience, C/ Faraday 9, Campus de Cantoblanco, 28049 Madrid, Spain}

\author{Matteo Jugovac}
\affiliation{Elettra Sincrotrone Trieste, Strada Statale 14 km 163.5, 34149 Trieste, Italy}

\author{Julien Rault} 
\affiliation{Synchrotron SOLEIL, Saint-Aubin, 91192 Gif-sur-Yvette, France}

\author{Patrick Le F\`evre} 
\affiliation{Synchrotron SOLEIL, Saint-Aubin, 91192 Gif-sur-Yvette, France}

\author{Fran\c{c}ois Bertran} 
\affiliation{Synchrotron SOLEIL, Saint-Aubin, 91192 Gif-sur-Yvette, France}

\author{Donya Mazhjoo} 
\affiliation{Peter Gr{\"{u}}nberg Institute and Institute for Advanced Simulation, \\
Forschungszentrum J{\"{u}}lich, D-52425 J{\"{u}}lich, Germany}

\author{Gustav Bihlmayer} 
\affiliation{Peter Gr{\"{u}}nberg Institute and Institute for Advanced Simulation, \\
Forschungszentrum J{\"{u}}lich, D-52425 J{\"{u}}lich, Germany}

\author{Stefan Bl\"ugel} 
\affiliation{Peter Gr{\"{u}}nberg Institute and Institute for Advanced Simulation, \\
Forschungszentrum J{\"{u}}lich, D-52425 J{\"{u}}lich, Germany}

\author{Rodolfo Miranda} 
\affiliation{IMDEA Nanoscience, C/ Faraday 9, Campus de Cantoblanco, 28049 Madrid, Spain}
\affiliation{Departamento de F\'isica de la Materia Condensada, Instituto Nicol\'as Cabrera and Condensed Matter \\ 
Physics Center (IFIMAC), Universidad Aut\'onoma de Madrid, Campus de Cantoblanco, 28049 Madrid, Spain}

\author{Julio Camarero} 
\affiliation{IMDEA Nanoscience, C/ Faraday 9, Campus de Cantoblanco, 28049 Madrid, Spain}
\affiliation{Departamento de F\'isica de la Materia Condensada, Instituto Nicol\'as Cabrera and Condensed Matter \\ 
Physics Center (IFIMAC), Universidad Aut\'onoma de Madrid, Campus de Cantoblanco, 28049 Madrid, Spain}

\author{Miguel Angel Valbuena}
\thanks{Authors to whom any correspondence should be addressed. E-mails: \href{miguelangel.valbuena@imdea.org}{miguelangel.valbuena@imdea.org}; \href{paolo.perna@imdea.org}{paolo.perna@imdea.org}}
\affiliation{IMDEA Nanoscience, C/ Faraday 9, Campus de Cantoblanco, 28049 Madrid, Spain}

\author{Paolo Perna}
\thanks{Authors to whom any correspondence should be addressed. E-mails: \href{miguelangel.valbuena@imdea.org}{miguelangel.valbuena@imdea.org}; \href{paolo.perna@imdea.org}{paolo.perna@imdea.org}}
\affiliation{IMDEA Nanoscience, C/ Faraday 9, Campus de Cantoblanco, 28049 Madrid, Spain}

\begin{abstract} 
Epitaxial graphene/ferromagnetic metal (Gr/FM) heterostructures deposited onto heavy metals (HM) have been proposed for the realization of novel spintronic devices because of their perpendicular magnetic anisotropy and sizeable Dzyaloshinskii-Moriya interaction (DMI), allowing for both enhanced thermal stability and stabilization of chiral spin textures. However, establishing routes towards this goal requires the fundamental understanding of the microscopic origin of their unusual properties. Here, we elucidate the nature of the induced spin-orbit coupling (SOC) at Gr/Co interfaces on Ir. Through spin- and angle-resolved photoemission along with density functional theory, we show that the interaction of the HM with the C atomic layer via hybridization with the FM is the source of strong SOC in the Gr layer. Furthermore, our studies on ultrathin Co films underneath Gr reveal an energy splitting of $\sim$\,100 meV (negligible) for in-plane (out-of-plane) spin polarized Gr $\uppi$ bands, consistent with a Rashba-SOC at the Gr/Co interface, which is either the fingerprint or the origin of the DMI. This mechanism vanishes at large Co thicknesses, where neither in-plane nor out-of-plane spin-orbit splitting is observed, indicating that Gr $\uppi$ states are electronically decoupled from the HM. The present findings are important for future applications of Gr-based heterostructures in spintronic devices.
\end{abstract}  
\maketitle

\section{INTRODUCTION}

Spintronics aims at exploiting the spin degree of freedom of electrons for new forms of information storage and logic devices \cite{Spintronic2015}. A major challenge for innovative, high-speed, low-power operation spintronic devices is to develop suitable spin transport channels with long spin lifetime and propagation length. Graphene (Gr) is an ideal spin channel material, exhibiting the longest spin relaxation length ever measured at room temperature, of several micrometers \cite{CastroNeto2009,Han2014,Avsar2019,Drogeler2016,Guimaraes2014,2015Ingla-Aynes}, as well as enhanced spin-to-charge conversion and spin Hall effect when interfaced with heavier atoms \cite{Yan2017,SaveroTorres2017}. This is due to the spin-orbit coupling (SOC), which is an important spin interaction that relates the spin to the momentum of electrons. While this interaction can drive to a modification of spin relaxation or spin diffusion lengths, strong SOC also dictates a vast variety of related intriguing phenomena, especially when reduced dimensions are at place. Prominent examples are the quantum spin Hall and anomalous Hall states \cite{Soumyanarayanan2016,Kane2005,Tse2011,Bernevig1757,Chang167,Weeks2011,Zhang2012}, skyrmions \cite{Fert2013,Yu2010}, as well as the interfacial Rashba and the Dzyaloshinskii-Moriya interaction (DMI), enabling the development of efficient non-volatile storage technologies \cite{Cai2017,Yu2014,Cao2020,Cao2022}. 

However, Gr is a material that presents a negligible SOC in its pristine state. At the \Kbar\ point of the Gr surface Brillouin zone (SBZ), where the Dirac point (DP) is located at the Fermi level (E$_{\textrm{F}}$), the SOC-induced gap is very small, of only 24-50 $\upmu$eV \cite{Gmitra2009}. To observe spin-orbit effects suitable for applications, it is thus necessary to enhance the SOC in Gr while preserving its Dirac cone structure. The high nuclear charge required can in principle be achieved by combining Gr with the appropriate substrates \cite{Marchenko2013,Avsar2014,Balakrishnan2014,Dedkov2008}, or by intercalation of heavy atoms in suitable geometries \cite{Marchenko2012,Otrokov2018,Calleja2015,Marchenko2015,Klimovskikh2015}, resulting in a strong hybridization and a spin-splitting, which can be deliberately induced in the vicinity of the DP. For instance, gold intercalation at the Gr/Ni interface has been demonstrated to create a giant spin-orbit splitting ($\sim$\,100 meV) of the Gr Dirac cone up to the Fermi energy \cite{Marchenko2012}, attributed to an in-plane Rashba spin polarization caused by hybridization with Au states. Besides, in the case of Gr/Pb bilayers grown on Pt(111) \cite{Calleja2015,Klimovskikh2017}, the observed enhancement of SOC has been explained in terms of intrinsic (Kane and Mele type \cite{Kane2005}) and extrinsic (Rashba \cite{Otrokov2018}) SOC contributions, responsible for out-of-plane and in-plane spin polarizations, respectively.

In this context, the intercalation of thin ferromagnetic metal (FM) layers underneath Gr is considered a promising approach for developing multifunctional devices with advanced capabilities as it enables unprecedented control of the spin filtering and injection efficiency \cite{Karpan2007, anadonAPL2021, Piquemal-Banci2020}, exchange coupling \cite{valvidaresnatComm2017}, tunnel magnetoresistance \cite{Cobas2012}, Rashba effect \cite{Otrokov2018, Rader2009, Varykhalov2008} and perpendicular magnetic anisotropy (PMA) \cite{Rougemaille2012,blancoACS2021, Yang2016, Yang2018, Ajejas2018,Vo-Van2010}, as well as the stabilization of magnetic skyrmions \cite{ollerosACS2020, Wang2020}. The combination of these properties opens up exciting possibilities for efficient electrical control of topologically-protected chiral spin textures in Gr/FM heterostructures, where an unexpectedly large interfacial DMI has been observed to play a key role \cite{Yang2018,Ajejas2018}. However, this raises questions about how Gr can induce a large interfacial DMI as it lacks strong SOC. While the strong hybridization between Gr and FM states \cite{Usachov2015,Pacile2014,Vita2014,JUGOVAC2020341} has been previously identified as one of the key ingredients underlying the enhancement of interfacial PMA \cite{blancoACS2021,Yang2016}, direct spectroscopic evidence of the exact microscopic mechanism that is ultimately responsible for the significant increase of DMI at Gr/FM interfaces remains elusive.

Earlier on, it has been suggested that Gr plays an essential role in determining the unusual properties of Gr/FM interfaces in the absence of strong SOC. For instance, on the one hand, in the case of Gr/Co interfaces, it has been experimentally found that the hybridization of Co 3$d$ orbitals with Gr bands can lead to an enhancement of the interfacial PMA \cite{blancoACS2021,Yang2016} via the anisotropy of the orbital magnetic moments \cite{blancoACS2021}. On the other hand, based on density functional theory (DFT) calculations of the electronic structure of Gr/Co bilayers on Ru(0001), the physical origin of the DMI at Gr/Co interfaces has been attributed to a Rashba effect originating from large variations of the SOC energy in the Co layer \cite{Yang2018}. These variations have been explained by the strong influence that the change in the potential gradient induced by Gr at the interface has on the Co 3$d$ orbitals \cite{Yang2018,Ajejas2018}. However, in this case, a Rashba splitting as small as 1.28 meV is predicted to be the reason for a considerable enhancement of the effective SOC value at the interface \cite{Yang2018}. This is in contrast to the case of Co/Pt interfaces, where the DMI has been explained by a Fert-Levy model \cite{Fert1980,Fert_msf1991}, according to which the DMI at the vacuum/FM interface should scale with the SOC in the material that is on the non-magnetic side of the interface underneath the FM layer. Conversely, very recent DFT calculations revealed that the DMI at both Gr/Co and Co/Pt interfaces may have a common physical origin, with the effect of Gr being to reduce the total DMI of the hererostructure by the inversion of the chirality of the vacuum/Co interfacial DMI \cite{Blanco-Rey2021}.

To resolve these issues, we experimentally investigate the electronic origin of the large DMI at Gr/FM interfaces. To this end, we use angle-resolved photoemission (ARPES) with and without spin resolution to systematically characterize the electronic structure of high quality Gr-based heterostructures with atomically flat interfaces and a homogeneous intercalated Co layer sandwiched between Gr and an epitaxial Ir(111) buffer layer. Unlike previous studies in which Gr/FM interfaces of high quality were only grown on metallic single-crystal substrates, our heterostructures consist of interfaces of the same excellent quality grown on either metallic or insulating commercial oxide substrates. The latter enable their suitability for real spintronic devices in which injected currents (electrical or spin) are not drained by the substrate \cite{blancoACS2021,Ajejas2018,Ajejas2020}. Our main finding is that a strong SOC induced in the Gr layer underlies the significant enhancement of interfacial DMI. The effect manifests itself by an in-plane energy spin-splitting of Gr $\uppi$-states which is consistent with a Rashba SOC at the Gr/Co interface two orders of magnitude larger than what was previously assumed. The experimental results are supported by DFT calculations pinpointing the tunable interaction between Gr and the heavy metal (HM) layer via the FM as the main source of SOC for C atoms. Our findings are important for the implementation of Gr-based heterostructures with advanced functionalities in future spintronic devices.

\section{RESULTS AND DISCUSSION}

\subsection {{\bf Electronic properties of pristine and Co-intercalated Gr/Ir/Al$_2$O$_3$ heterostructures}}

\begin{figure}
\centering
\includegraphics[width=0.43\textwidth]{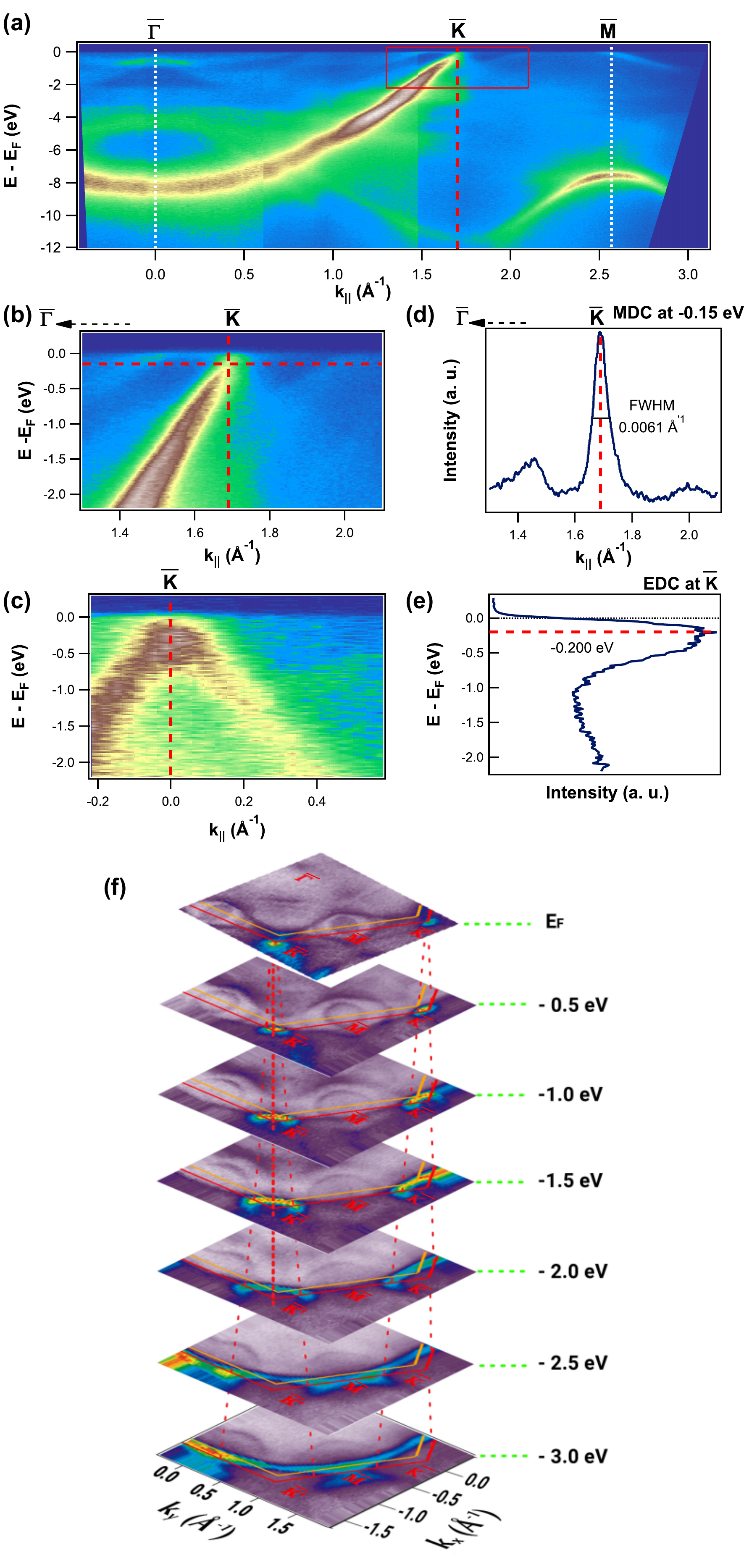}
\caption{{\bf Electronic structure of Gr/Ir/Al$_{\textbf{2}}$O$_{\textbf{3}}$(0001)}. ARPES intensity band map along the $\overline{\Gamma}$-$\overline{\mbox{K}}$-$\overline{\mbox{M}}$ direction. (b) Zoom-in over the region of the Dirac cone corresponding to the red rectangular area in (a). (c) Perpendicular cut to the $\overline{\Gamma}$-$\overline{\mbox{K}}$ direction centered at the \Kbar\ point. (d) MDC and (e) EDC profiles extracted at the energy and momentum positions indicated in (b) and (c), respectively. (f) FS and CE maps. Red dashed vertical lines indicate the Gr Dirac cones at \Kbar\ and \Kbarprime. Ir and Gr SBZs are indicated by orange and red hexagons, respectively.}
\label{Fig1_ARPES_Gr_Ir}
\end{figure}

\begin{figure}[t!]
\centering
\includegraphics[width=0.48\textwidth]{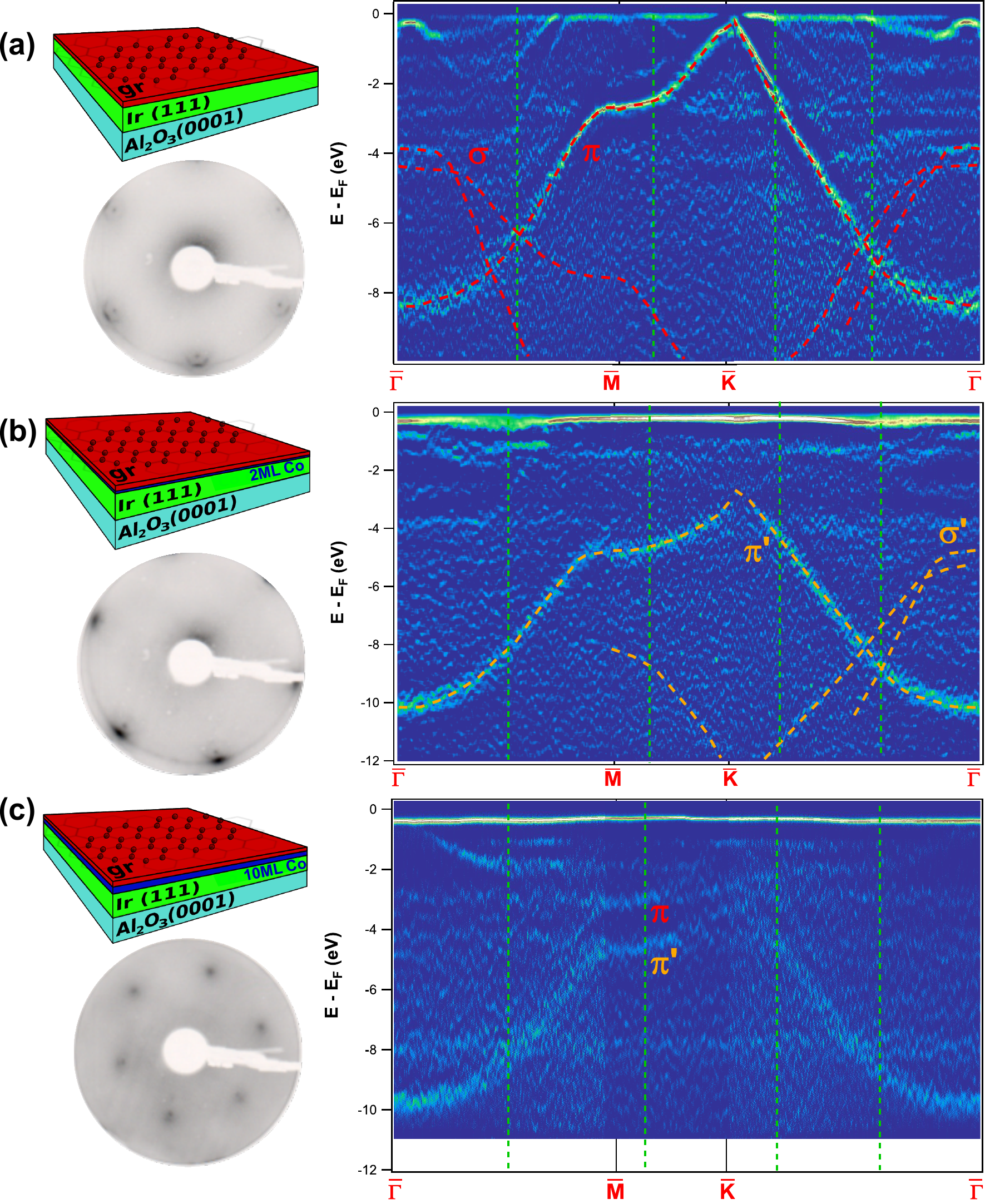}
\caption{{\bf Characterization of Co-intercalated 
Gr/Ir/Al$_{\textbf{2}}$O$_{\textbf{3}}$(0001) heterostructures}. Evolution of the electronic properties. (a)-(c) LEED images and curvature ARPES band maps. The band dispersions provide an overview of the electronic structure of (a) Gr/Ir and Co-intercalated (b) Gr\,/2\,ML\,Co\,/Ir and (c) Gr\,/10\,ML\,Co\,/Ir systems along the
$\overline{\Gamma}$-$\overline{\mbox{M}}$-$\overline{\mbox{K}}$-$\overline{\Gamma}$
direction. Red and orange dashed lines follow the Gr $\uppi$ and $\upsigma$ bands for Gr/Ir and Gr\,/2\,ML\,Co\,/Ir, respectively. On the left, the schematic sketches of the corresponding heterostructures and their LEED patterns acquired at energies of (a),(b) 69 eV and (c) 155 eV are shown. Note that the ARPES band maps were obtained after applying the curvature algorithm of Ref.~\onlinecite{Zhang2011} in order to obtain sharper and better resolved electronic band dispersions (raw data shown in Fig.~S1 in SI). Green dashed vertical lines in the ARPES band maps indicate the positions at which the EDC profiles shown in supplementary Fig.~S2 have been extracted.} 
\label{Fig2_bands_comparison}
\end{figure}

In order to understand the importance of the hybridization effects involving Gr $\uppi$-states for the enhancement of SOC at the interface, we first performed ARPES measurements of the electronic structure before and after Co intercalation. The procedure adopted for the preparation of epitaxial Gr-based heterostructures is described in more detail in the Methods section. A 30-nm-thick Ir buffer was DC sputtered at 670~K on an insulating Al$_2$O$_3$(0001) substrate, onto which epitaxial Gr was subsequently grown by ethylene dissociation at 1025~K. After Gr growth, Co was deposited  by electron-beam evaporation at room temperature, while its intercalation under Gr was promoted by a moderate thermal annealing. The high quality of the interfaces and the completion of the intercalation process were checked by ARPES and low-energy electron diffraction (LEED), see Figs.~\ref{Fig1_ARPES_Gr_Ir} and \ref{Fig2_bands_comparison}. The Gr/Ir(111) structure was demonstrated to exhibit the well-known $\sim$10$\times$10 moir\'e pattern due to the coexistence of the incommensurate unit cells of Ir and Gr \cite{blancoACS2021, Ajejas2020, NDiaye2008}, as can be seen in Fig.~\ref{Fig2_bands_comparison}(a).

The energy-momentum band dispersion of Gr/Ir/Al$_2$O$_3$ (0001) measured by ARPES along the $\overline{\Gamma}$-$\overline{\mbox{K}}$-$\overline{\mbox{M}}$ direction of the Gr SBZ is depicted in Figs.~\ref{Fig1_ARPES_Gr_Ir}(a)-\ref{Fig1_ARPES_Gr_Ir}(c). The overall electronic structure is consistent with the characteristic quasi-freestanding character of the Gr layer on Ir(111), and comparable to the case of Gr grown on Ir(111) single-crystal substrates \cite{Klimovskikh2015, Pletikosic2009, Kralj2011, Pletikosic2010}, proving the high quality of the interface. The measured Gr $\upsigma$ and $\uppi$ bands are very sharp and intense, as evidenced by the momentum-distribution curve (MDC) at an energy of E-E$_{\textrm{F}}$\,=\,-150~meV and the energy-distribution curve (EDC) at the \Kbar\ point of the Gr SBZ (Figs.~\ref{Fig1_ARPES_Gr_Ir}(d) and \ref{Fig1_ARPES_Gr_Ir}(e), respectively), confirming again the interface quality. The dominant feature in the spectrum is the Gr $\uppi$ band. The bottom of the $\uppi$ band at \pntG\ is located at an energy of E-E$_{\textrm{F}}$\,$\approx$\,-8.28 eV and disperses linearly towards the E$_{\textrm{F}}$, resulting in the formation of a slightly $n$-doped Dirac cone, as inferred from the MDC and EDC analyses. The narrowest MDC corresponding to an energy of -150~meV can be fitted to a Lorentzian function, whose full width at half maximum (FWHM) is $\approx$\,0.0061\,\invA, similar to the FWHM value obtained for Gr grown on Ir(111) single-crystal substrates \cite{Kralj2011}. In the band dispersion the $\uppi$ band replicas can be distinguished, which appear due to the additional periodicity of the moir\'e lattice despite their much lower intensity when compared to the main Gr Dirac cone at this photon energy (h$\upnu$\,=\,64~eV).

The Dirac cones are also visible at the Fermi surface (FS) and constant energy (CE) maps shown in Fig.~\ref{Fig1_ARPES_Gr_Ir}(f). Here, the main Dirac cones at \Kbar\ and \Kbarprime\ points, whose dispersion is indicated by red dashed lines, exhibit the characteristic trigonal symmetry of quasi-freestanding Gr on Ir(111), and coexist with the band dispersion of Ir 5$d$ states typically observed in Ir(111) bulk single crystals \cite{Vita2014,Starodub2011}. Note that these main Ir 5$d$ ellipsoidal-shaped features close to the E$_{\textrm{F}}$ are also visible in the curvature ARPES band maps of Fig.~\ref{Fig2_bands_comparison}(a). The Gr and Ir SBZs, indicated by red and orange hexagons in Fig.~\ref{Fig1_ARPES_Gr_Ir}(f), can be clearly identified from the intensity of the CE contours associated to the different features. Additionally, hybridization between Ir 5$d$ and Gr $\uppi$ states can be observed in the energy region between -1 and -4 eV, resulting into a reduction of the spectral weight at the band crossings as seen along the $\overline{\Gamma}$-$\overline{\mbox{K}}$ direction. Nonetheless, the hybridization effects are sufficiently weak that the overall quasi-relativistic Dirac-like dispersion is largely preserved, as in the case of weakly bonded systems, where the electronic structure is usually similar or close to that of freestanding Gr.
 
\begin{figure*}
\centering
\includegraphics[width=0.8\textwidth]{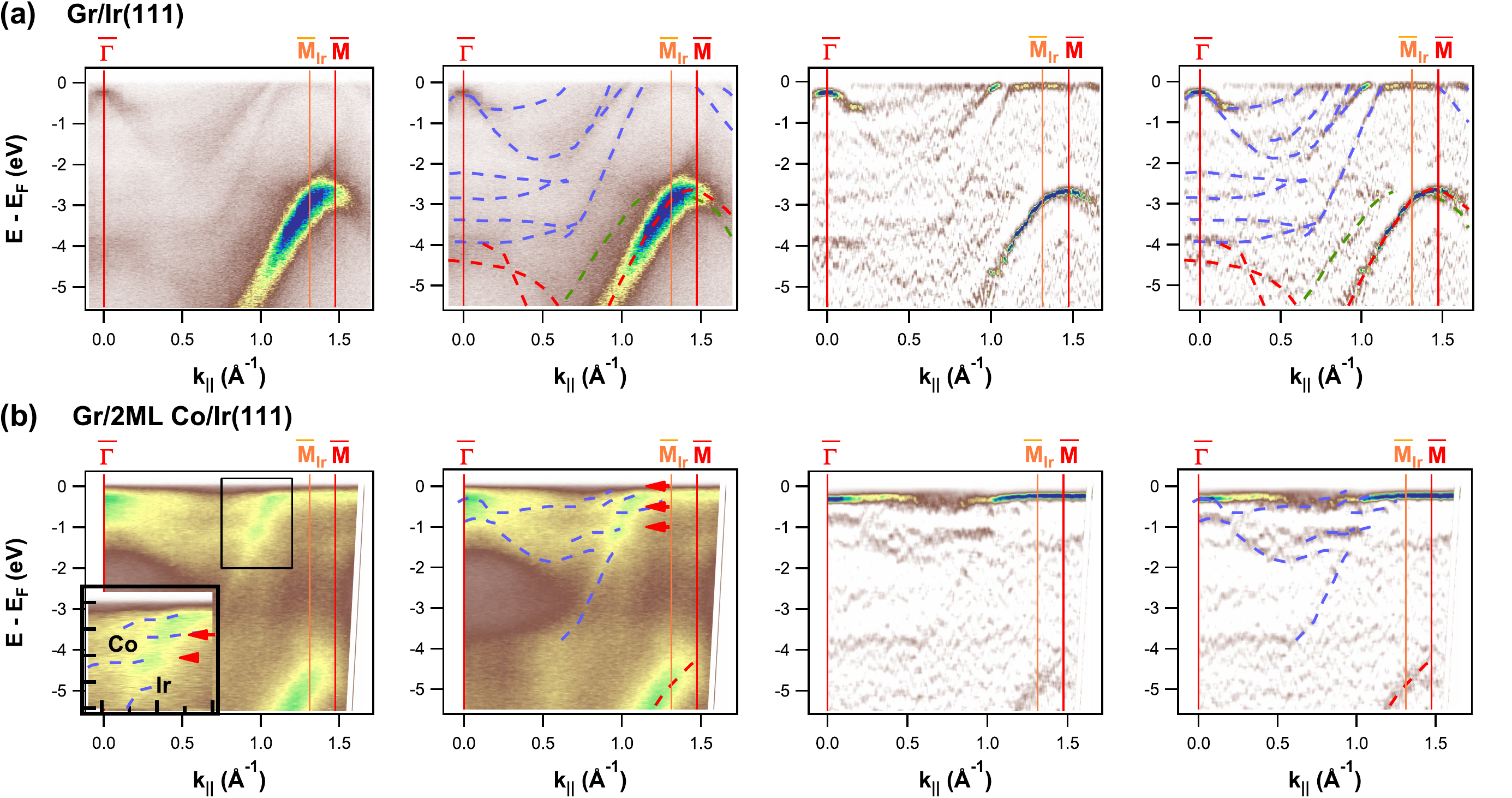}
\caption{{\bf Hybridization effects in the electronic structure}. Zoom-in on the dispersion of the bands over the region close to the E$_{\textrm{F}}$ for (a) Gr/Ir and (b) Gr\,/2\,ML\,Co\,/Ir heterostructures along the $\overline{\Gamma}$-$\overline{\mbox{M}}$ direction. Raw and curvature ARPES band maps with and without superimposed guidelines are presented, respectively. Blue dashed lines indicate the Ir-derived bands (a), which persist and hybridize with Co states (b), as evidenced by the corresponding gap openings. A close view of the gap regions is shown as an inset on the left panel in (b). Red and green dashed lines follow the Gr bands and a Gr replica originating from the moir\'e superlattice, respectively. Red arrows indicate the energies at which the CE contours shown in Fig.~\ref{Fig4_FS_CE} have been obtained.}
\label{Fig3_GM_hyb}
\end{figure*}

\begin{figure*}[!]
\centering
\includegraphics[width=0.85\textwidth]{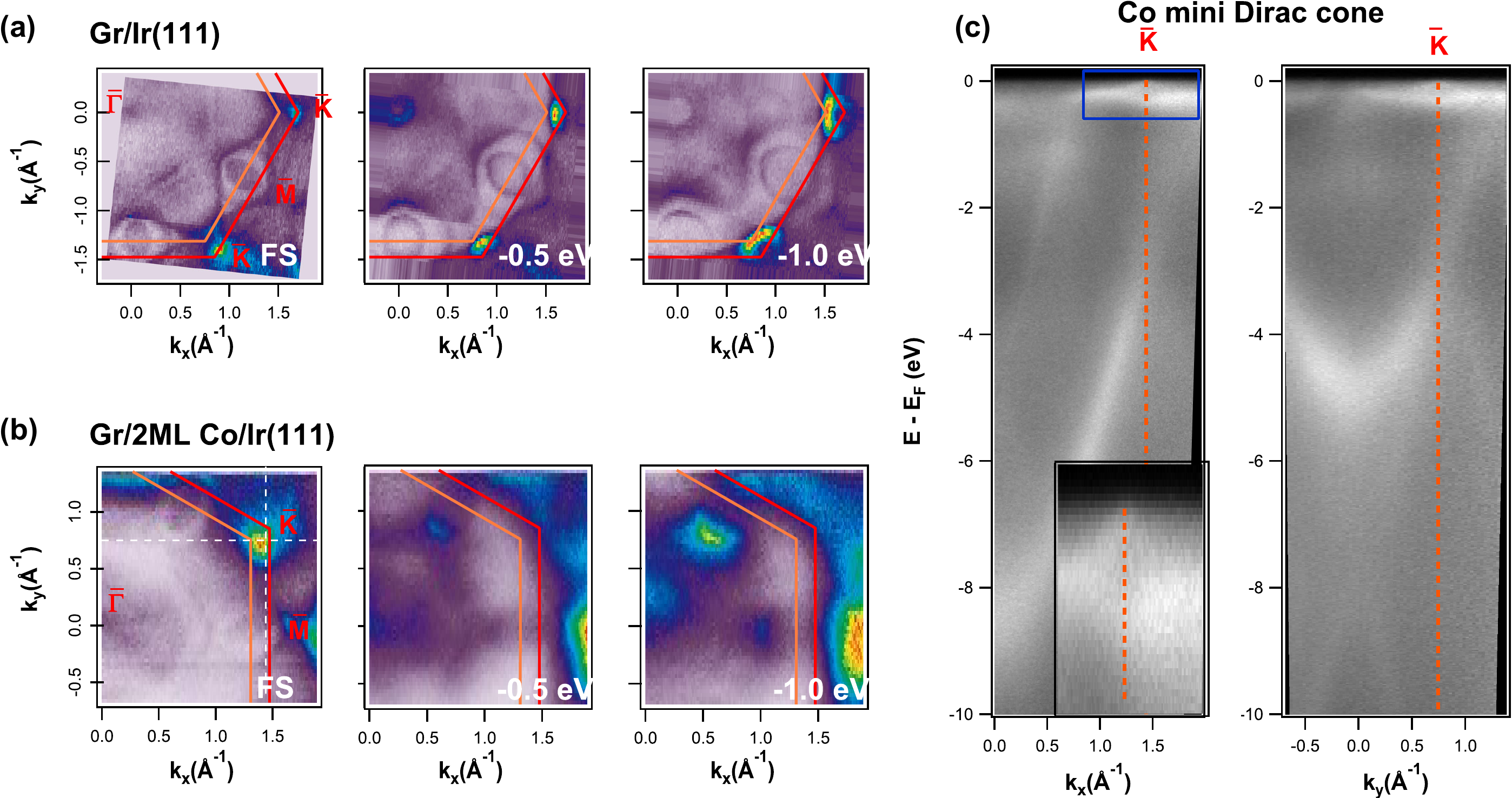}
\caption{{\bf Evolution of the electronic structure in momentum space}.~(a),(b)~Fermi surface (FS) and constant energy contours acquired at energies of -0.5 and -1.0 eV for (a) Gr/Ir and (b) Gr\,/2\,ML\,Co\,/Ir heterostructures. In each panel, the Gr and Ir hexagonal SBZs are indicated by red and orange solid lines, respectively. (c) Development of mini Dirac cones following Co intercalation. The band dispersions were taken along the two orthogonal momentum directions indicated by the white dashed lines crossing at \Kbar\ point in (b). On the left, a zoom-in over the region of the mini Dirac cone corresponding to the blue rectangular area near \Ef\ is shown as an inset.}
\label{Fig4_FS_CE}
\end{figure*}

In Fig.~\ref{Fig2_bands_comparison} and Fig.~S1 in Supplementary Information (SI), we provide an overview of the electronic band structure as seen by ARPES along the $\overline{\Gamma}$-$\overline{\mbox{M}}$-$\overline{\mbox{K}}$-$\overline{\Gamma}$ direction before (Fig.~\ref{Fig2_bands_comparison}(a)) and after intercalation of an epitaxial Co film of 2 and 10 monolayers (ML) thickness (Figs.~\ref{Fig2_bands_comparison}(b) and~\ref{Fig2_bands_comparison}(c), respectively). The LEED images in Figs.~\ref{Fig2_bands_comparison}(a) and~\ref{Fig2_bands_comparison}(b) display signatures of a moir\'e pattern and a clear commensurate Gr structure can be observed in the corresponding panel of Fig.~\ref{Fig2_bands_comparison}(c), indicating in all cases a complete Co intercalation. In-line with previous discussion, the overall energy-momentum dispersion of Gr $\uppi$ and $\upsigma$ states is close to that of a weakly interacting layer system (Fig.~\ref{Fig2_bands_comparison}(a)). Similar to Fig.~\ref{Fig1_ARPES_Gr_Ir}, here one can recognize between -4~eV and the E$_{\textrm{F}}$ the dispersing Ir-derived bands which, as also observed in Figs.~\ref{Fig3_GM_hyb}(a) and \ref{Fig4_FS_CE}(a), consist mainly of the two concentrical ellipsoidal pockets centered at the $\overline{\mbox{M}}_{\rm Ir}$ point of Ir SBZ \cite{Vita2014,Starodub2011} and the Rashba-split Ir surface state (SS) dispersing towards $\overline{\Gamma}$ \cite{Varykhalov2012}. 

Upon Co intercalation, the overall electronic band structure of Gr/Ir becomes highly modified because of the strong interaction with the Co layer. As a result, the Gr $\uppi$ band for a 2-ML-thick intercalated Co film is already significantly shifted down in energy by $\sim$\,2~eV, as seen in Fig.~\ref{Fig2_bands_comparison}(b) and Figs.~S1 and S2 in SI. In particular, the bottom of the $\uppi$ band at the $\overline{\Gamma}$ point is shifted from -8.28 to -10.08 eV, and the Dirac cone merges with Co 3$d$ states at an energy of $\sim$\,-2.80 eV in the vicinity of the $\overline{\mbox{K}}$ point. Despite the fact that the intensity and sharpness of the Gr bands are reduced when compared to the case of Gr/Ir in Fig.~\ref{Fig2_bands_comparison}(a), the resulting changes in the overall band dispersion can still be clearly observed. Further Co intercalation results in a double-like $\uppi$ band dispersion as seen in the ARPES maps of Fig.~\ref{Fig2_bands_comparison}(c) and in the EDC profiles extracted at the momentum positions indicated by green dashed lines in Fig.~\ref{Fig2_bands_comparison} and reported in Fig.~S2 (SI). 
The double $\uppi$ band feature can be interpreted as a nearly-undisturbed Gr/Ir $\uppi$ band closer to the E$_{\textrm{F}}$ plus the more electronically doped $\uppi^{\prime}$ Gr/Co at higher binding energies, associated with a highly-interacting, commensurate phase. As reported earlier \cite{blancoACS2021}, the intercalation of further Co layers leads to a relaxation of the Co lattice resulting in a strong corrugation of the Gr layer. Although the intercalation of Co underneath Gr is layer-by-layer and epitaxial \cite{blancoACS2021}, the formation of Co clusters or inhomogeneities at large thicknesses cannot be completely ruled out and can eventually lead to a more corrugated Gr layer \cite{blancoACS2021, Pacile2014}. The double-like $\uppi$ band dispersion can thus be associated to a corrugated Gr layer where the $\uppi$ band feature would appear due to weakly interacting, hill, C atoms, while the $\uppi^{\prime}$ band would arise from highly interacting, valley, C atoms. As indicated by the EDC profiles shown in Fig.~S2 (SI), this doubly-dispersing state is consistent with a simultaneous contribution from both highly interacting Gr-Co $\uppi^{\prime}$ states and a less interacting Gr-Ir $\uppi$ band (blue and red dashed lines in Fig. S2), respectively), in qualitative agreement with previous findings on other intercalated systems \cite{Pacile2014}.

Furthermore, in Figs.~\ref{Fig2_bands_comparison}(b) and \ref{Fig2_bands_comparison}(c) several prominent Co electronic states can be identified: (i) a narrow and highly intense Co peak close to the E$_{\textrm{F}}$, present along the entire $\overline{\Gamma}$-$\overline{\mbox{M}}$-$\overline{\mbox{K}}$-$\overline{\Gamma}$
direction in an energy range between -0.2 and -0.5 eV, and common to the two selected Co thicknesses; (ii) for the thicker Co system (Fig. \ref{Fig2_bands_comparison}(c)), a highly localized and broad state located at an energy of -8.0 eV is present in the band structure, which could be indicative of a more disordered thicker, though epitaxial, Co layer; (iii) another Co-derived band strongly dispersing from -1.9 eV to the E$_{\textrm{F}}$ towards $\overline{\Gamma}$, which confirms that, aside from the possible disorder, the intercalation of the thickest Co layer is mainly epitaxial; (iv) other localized states can be observed for the 10 ML Co system at -1.1 and -2.9 eV, while Ir features disappear as the Co thickness is increased.

Let us now turn our attention to the electronic properties of the 2\,ML\,Co system in more detail. Previous ARPES measurements have shown how after intercalation of one or more Co layers, the Ir bands close to the E$_{\textrm{F}}$ can barely be distinguished, while the Co 3$d$ states are so intense and broad that the Ir 5$d$ bands become blurred and diffuse \cite{Pacile2014,Vita2014}. A zoom-in on the electronic structure from the E$_{\textrm{F}}$ to -5.5 eV for both Gr/Ir and Gr\,/2\,ML\,Co\,/Ir systems along the $\overline{\Gamma}$-$\overline{\mbox{M}}$ and $\overline{\Gamma}$-$\overline{\mbox{K}}$ directions is shown in Fig.~\ref{Fig3_GM_hyb} and Fig.~S3 (SI), respectively. Raw and curvature ARPES band maps are displayed for both systems. Correspondingly, the FS and CE maps are also presented in Figs.~\ref{Fig4_FS_CE}(a) and \ref{Fig4_FS_CE}(b), as well as Fig. S4 in SI.

First, it should be noted that, contrary to previous studies, different bands corresponding to the electronic states of the underlying Ir(111) layer can always be detected. Noticeably, the electronic states that make up the ellipsoidal electron pockets centered at $\overline{\mbox{M}}_{\rm Ir}$ (Fig.~\ref{Fig4_FS_CE}(a)) are unambiguously resolved at the FS (Fig.~\ref{Fig4_FS_CE}(b)), namely the prominent Ir-derived band dispersing from -3.8 eV at 0.6 \invA\ to the E$_{\textrm{F}}$ at 1.13 \invA\ along the $\overline{\Gamma}$-$\overline{\mbox{M}}$ direction (Fig.~\ref{Fig3_GM_hyb}(b)), which coincides with the same band for the Gr/Ir system (Fig.~\ref{Fig3_GM_hyb}(a)). Particularly, hybridization between Co 3$d$ states at $\sim$\,-0.5 and -1.2 eV (clearly resolved between 0.5 and 0.8 \invA\ along the $\overline{\Gamma}$-$\overline{\mbox{M}}$) direction and the Ir dispersing band becomes evident, as reflected in the regions of increased intensity (red arrows on raw ARPES band maps in Fig.~\ref{Fig3_GM_hyb}(b), left) and by the gap opening observed in both, raw and curvature, ARPES band maps (blue dashed lines in Fig.~\ref{Fig3_GM_hyb}(b), right). The main gap opening of Ir dispersing band occurs between -1.1 and -0.65 eV at 0.99 \invA\ along 
 the $\overline{\Gamma}$-$\overline{\mbox{M}}$ direction(magnified in the inset of Fig.~\ref{Fig3_GM_hyb}(b)). 

This hybridization is also reflected in the CE maps at -0.5 and -1.0 eV in Figs.~\ref{Fig4_FS_CE}(a) and \ref{Fig4_FS_CE}(b): the Ir ellipsoidal pockets near the E$_{\textrm{F}}$ in Gr/Ir (Fig.~\ref{Fig4_FS_CE}(a)), become closed ellipses when interacting with these Co 3$d$ states (Fig.~\ref{Fig4_FS_CE}(b)). Note also that, besides the Co-Ir hybridization close to $\overline{\mbox{M}}_{\rm Ir}$, the Ir Rashba SS \cite{Varykhalov2012}, with negative effective mass and maximum intensity at -0.26 eV at $\overline{\Gamma}$, is still preserved after Co intercalation (albeit slightly modified by the interaction with Co 3$d$ states), as it can be inferred from the band maps along both  $\overline{\Gamma}$-$\overline{\mbox{M}}$ (Fig.~\ref{Fig3_GM_hyb}) and $\overline{\Gamma}$-$\overline{\mbox{K}}$ (Fig.~S3 in SI) directions. Therefore, the Ir Rashba SS at $\overline{\Gamma}$ and Ir 5$d$ bands close to $\overline{\mbox{M}}$ are preserved and hybridized after the intercalation of 2\,ML\,Co. Another significant feature of the electronic structure of the 2\,ML\,Co system is the development of mini Dirac cones in a narrow energy region of $\sim$\,0.2 eV below the E$_{\textrm{F}}$ at the $\overline{\mbox{K}}$ point (Fig.~\ref{Fig4_FS_CE}(c)). These mini Dirac cones have been previously observed in a perfectly oriented Gr layer on a Co(0001) surface and they strongly depend on the epitaxial quality of the interface \cite{Usachov2015}. The mini Dirac cones are due to the interaction of C~2$p_z$ and Co~3$d$ orbitals, have a fully two-dimensional character with their wave functions located at the interface, and are spin polarized. The development of the mini Dirac cone in the measured band dispersions is another hallmark of both the Co-C interaction and our interface quality.

To summarize this part, after 2\,ML of intercalated Co, the SS and Ir 5$d$ bands are preserved at the E$_{\textrm{F}}$ and become strongly modified just below it due to the interaction with Co 3$d$ electronic states. Moreover, the perfectly oriented epitaxial layer leads to the formation of the mini Dirac cone as a result of the Co-C interaction. Further Co intercalation decouples Ir(111) electronic structure from the Gr/Co interface, resulting in the possibility of introducing some disorder effects. From our ARPES results we have provided direct spectroscopic evidence of the Co-Ir electronic hybridization, including the Ir(111) Rashba SS, as well as the Co-C electronic interaction. Thus, based on our findings, we can assign the Ir(111) layer interaction with the C atoms via the intercalated Co layer (at least for the lower Co thickness) as the main source of SOC at the Gr/Co interface. We note, however, that while this mechanism can result in high DMI values at Gr/Co interfaces, it also induces a strong SOC in the Gr layer which has previously been thought to be negligible and, therefore, not the reason for the observed enhancement of interfacial DMI \cite{Yang2018,Ajejas2018,Blanco-Rey2021}. 

\subsection {{\bf Large Rashba-type spin-orbit splitting of Gr $\uppi$-states at the Gr/Co interface}}

\begin{figure*}
\centering
\includegraphics[width=0.8\textwidth]{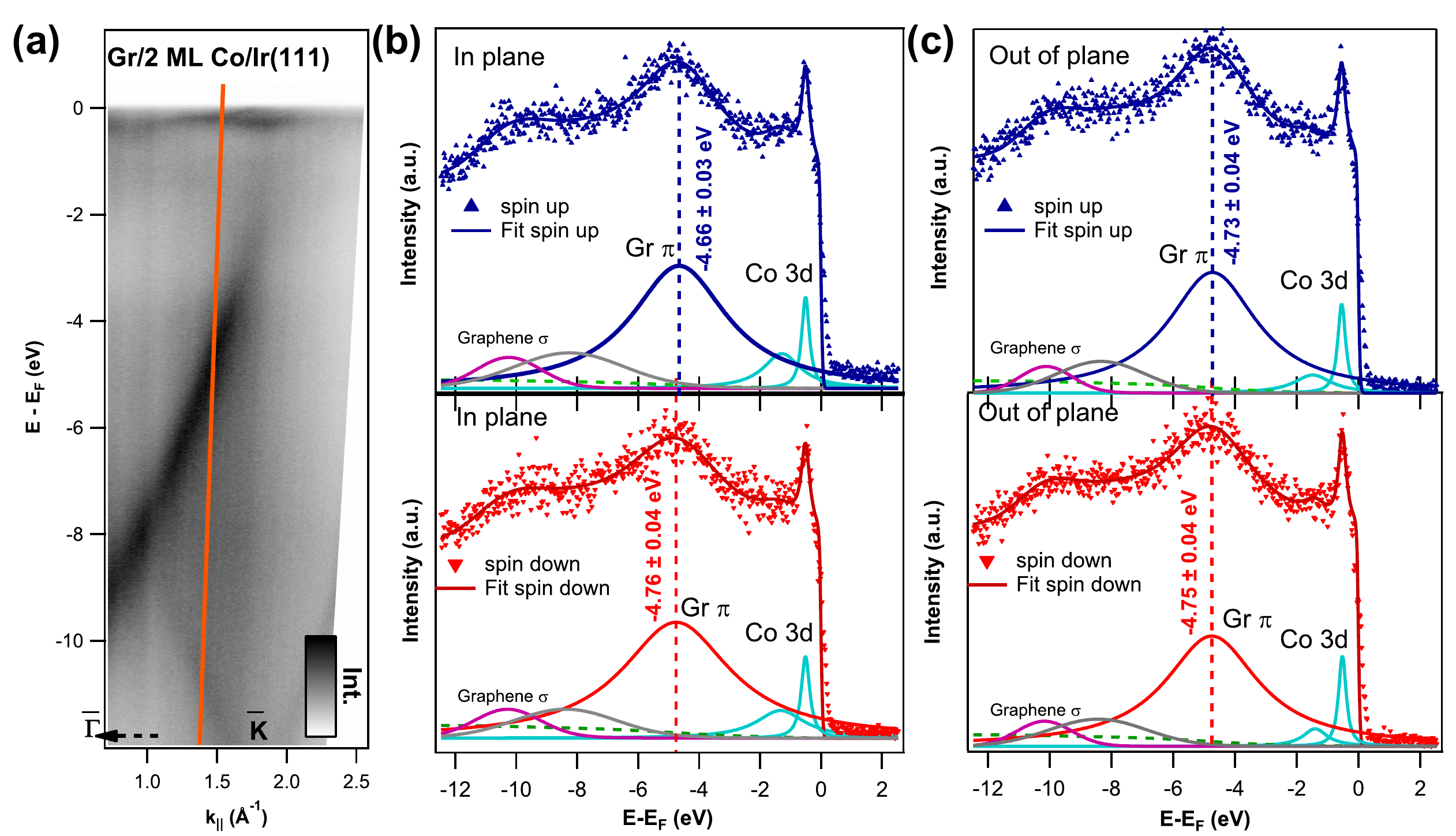}
\caption{{\bf Influence of hybridization on the spin structure of valence band states}. (a) Spin-integrated band dispersion of Gr\,/2\,ML\,Co\,/Ir taken along the $\overline{\Gamma}$-$\overline{\mbox{K}}$ direction using 64~eV photons. (b),(c) Normalized spin-resolved EDCs (blue/red colors for spin up/down respectively) corresponding to (b) the in-plane chiral and (c) out-of-plane spin
components, taken at the momentum position indicated by the orange vertical line in (a). Red and blue solid lines are fits to the experimental data. The in-plane spin projections are perpendicular to the momentum direction in (a). Individual peaks labelled according to the corresponding band features are displayed at the bottom of each panel.}
\label{Fig5_SR_ARPES_fits}
\end{figure*}

In order to investigate whether an extrinsic or intrinsic spin-orbit induced spin-splitting is generated in the Gr $\uppi$ states by Co intercalation, we performed spin-resolved ARPES measurements and DFT calculations. To this end, spin-resolved ARPES spectra were acquired at selected wave vectors, $k$, of the Gr SBZ 
(Fig\textcolor{red}{s}.~\ref{Fig5_SR_ARPES_fits}(a)). The spin-resolved EDCs for 
Gr\,/2\,ML\,Co\,/Ir\,/Al$_2$O$_3$(0001) are presented in Figs. \ref{Fig5_SR_ARPES_fits}(b) and \ref{Fig5_SR_ARPES_fits}(c) (in-plane chiral and out-of-plane spin components, respectively), measured at 1.48 \invA\ along the $\overline{\Gamma}$-$\overline{\mbox{K}}$ direction, as indicated by the dashed vertical line in the ARPES band map of Fig. \ref{Fig5_SR_ARPES_fits}(a), which crosses the region of the doped main Dirac cone with a stronger Gr character.

To isolate and extract the precise contribution from the Gr $\uppi$ band in each spin channel separately, the spin-resolved EDCs in Fig.~\ref{Fig5_SR_ARPES_fits} have been normalized and fitted using five different components, which correspond to: the two peaks closer to the E$_{\textrm{F}}$ arising from Co 3$d$ states (turquoise in Figs.~\ref{Fig5_SR_ARPES_fits}(b) and \ref{Fig5_SR_ARPES_fits}(c); the Gr $\upsigma$ band contribution (purple); the wider feature appearing after Co intercalation (grey, almost negligible contribution from disordered or localized states); and the intense contribution from Gr $\uppi$ states (blue for spin up, and red for spin down). This analysis has been applied consistently to both spin up and spin down spectra (blue and red colors respectively), which were simultaneously acquired using independent electron counters. 

\begin{figure*}
\centering
\includegraphics[width=0.85\textwidth]{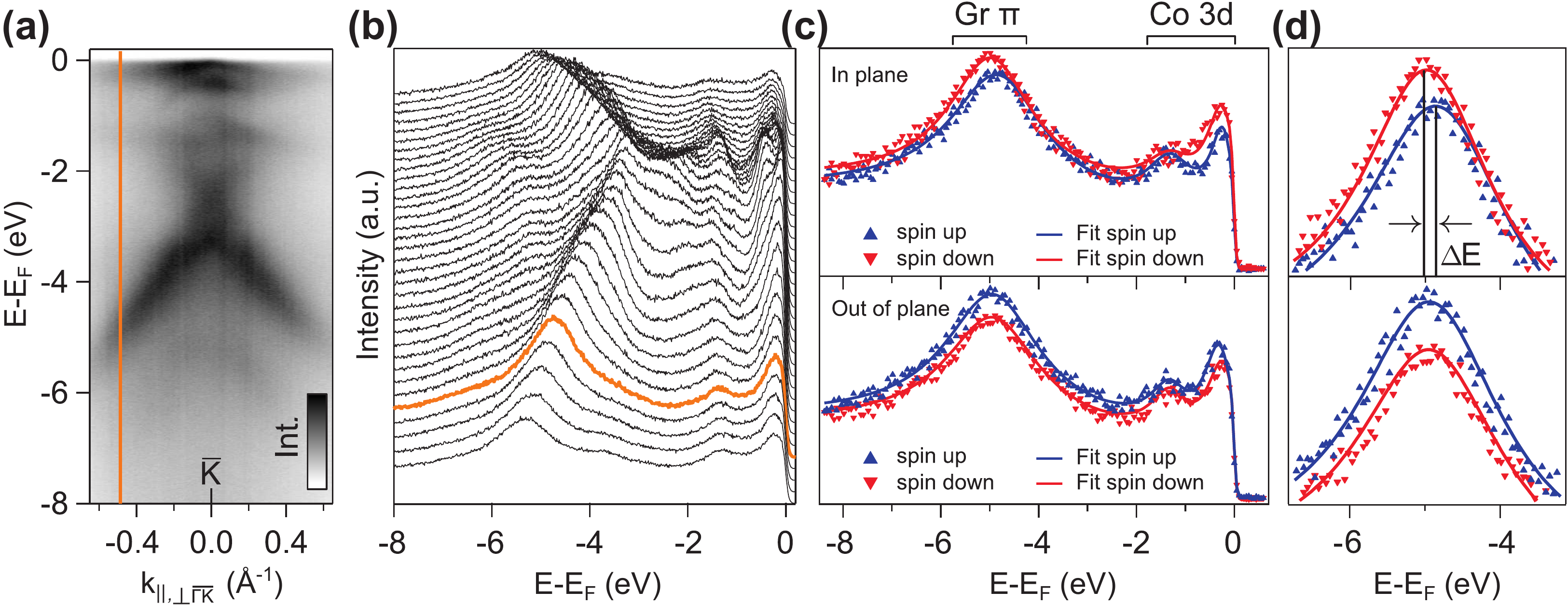}
\caption{{\bf Visualizing the spin structure of valence band states}. (a) ARPES band dispersion of Gr\,/4\,ML\,Co\,/Ir centered at the \Kbar\ point, taken at 45~eV photon energy and along the momentum direction of the Gr SBZ orthogonal to $\overline{\Gamma}$-$\overline{\mbox{K}}$. (b) Evenly-spaced EDC profiles over the full momentum range shown in (a). (c),(d) Spin-resolved EDCs taken at the momentum position indicated by both the vertical solid line in (a) and the spin-integrated EDC highlighted in (b). The spin-resolved EDCs corresponding to the in-plane chiral (perpendicular to momentum direction in (a)) and out-of-plane spin components are shown in the top and bottom panels respectively. Solid lines are fits to the data (see Methods). (d) Zoom-in over the range of the Gr $\uppi$ band. The spin splitting of the in-plane channel ($\Delta$E) is indicated.}
\label{Fig6_Bessy}
\end{figure*}

As clearly seen in Fig.~\ref{Fig5_SR_ARPES_fits}(b), there is a large energy spin-splitting in the in-plane chiral spin component for Gr $\uppi$ states of $\Delta$E$_{\textrm{\,in-plane}}$\,=\,(100\,$\pm$\,40)~meV. Conversely, the out-of-plane spin up and down spectra in Fig.~\ref{Fig5_SR_ARPES_fits}(c) do not show any sizeable spin-splitting above the average error of the curve fitting procedure, $\Delta$E$_{\textrm{\,out-of-plane}}$\,=\,(20\,$\pm$\,40)~meV. 
The energy spin-splitting seen in the in-plane chiral spin component for the Gr $\uppi$ band is consistent with the one reported for Au and Pb intercalated layers at similar $k$ values along the $\overline{\Gamma}$-$\overline{\mbox{K}}$ direction \cite{Marchenko2012,Otrokov2018}. In Fig.~\ref{Fig6_Bessy}, we further corroborate this result for 4 ML Co intercalated Gr on Ir(111) along the momentum direction of the Gr SBZ orthogonal to the $\overline{\Gamma}$-$\overline{\mbox{K}}$ direction. Note that similar to Fig.~\ref{Fig5_SR_ARPES_fits}, the in-plane chiral spin up and spin down orientations in Fig.~\ref{Fig6_Bessy}, which are perpendicular to the electron momentum, are also tangential to the Dirac cone, however rotated by 90 degrees with respect to Fig.~\ref{Fig5_SR_ARPES_fits}. 

In Figs.~\ref{Fig6_Bessy}(a) and \ref{Fig6_Bessy}(b), a spin-integrated band dispersion of the Dirac cone alongside spin-integrated EDCs are shown, taken with 45~eV photons and along the direction perpendicular to $\overline{\Gamma}$-$\overline{\mbox{K}}$ for a Gr/4 ML Co grown on Ir(111) single-crystal. Figure~\ref{Fig6_Bessy}(c) shows spin-resolved EDCs corresponding to the in-plane chiral and out-of-plane spin components, measured at the momentum position indicated by both the vertical solid line in Fig.~\ref{Fig6_Bessy}(a) and the spin-integrated EDC highlighted in Fig.~\ref{Fig6_Bessy}(b). As it can be also seen in a zoom-in over the range of the Gr $\uppi$ band in Fig.~\ref{Fig6_Bessy}(d), a large energy spin-splitting consistent with the spin-resolved measurements of Fig.~\ref{Fig5_SR_ARPES_fits} is clearly resolved in the in-plane chiral spin component for Gr $\uppi$ states. Taking altogether, the chiral spin up and spin down orientations in Fig.~\ref{Fig5_SR_ARPES_fits}(b) and \ref{Fig6_Bessy}(d) are consistent with a spin-orbit splitting predominantly associated with an in-plane Rashba-type spin texture that is characterized by a counterclockwise (clockwise) spin circulation around $\overline{\mbox{K}}$ for the outer (inner) $\uppi$ band spin states. This observation rules out the possibility of a collinear alignment of antiparallel spin vectors in a global spin texture whose origin would be entirely magnetic. In this respect, it should be noted that due to the energy dependent hybridization, spin-orbit effects with reducing energy are increasingly more difficult to resolve, as the lower half of the Dirac cone possesses very strong Co character, resulting in an overall spin texture that is more complex. This effect is manifested by an out-of-plane canting of electron spins which can be nevertheless observed in Fig.~\ref{Fig6_Bessy}(d), despite the reduced Co weight in this energy region. The energy spin-splitting $\Delta E$ is consistent with the previous estimation from  Fig.~\ref{Fig5_SR_ARPES_fits}.

Keeping this in mind, as next step, spin-resolved ARPES measurements were also performed on the thicker, 10 ML Co system (see Fig.~S5 in SI). Despite the increasing spectral broadening of Gr $\uppi$ states induced by Co intercalation, we do not resolve any spin-orbit splitting of considerable size between opposite spin states. This is consistent with our analysis of the Gr $\uppi$ band contribution in each spin channel, indicating that any possible energy spin-splitting is well below the fitting errors for both in-plane chiral and out-of-plane spin components. Such absence of measurable spin-orbit splitting for Gr $\uppi$ states in thicker intercalated Co films is in agreement with previous findings for Gr grown on Co bulk single-crystal substrates \cite{Rader2009}. 
These observations support our ARPES findings that at larger Co thickness the Gr/Co interface is electronically decoupled from the Ir underlayer, and emphasizes the critical role of the tunable interaction between Gr and the HM layer via the FM as source of SOC for C atoms. Accordingly, a large Rashba-SOC at the Gr/Co interface originating from the Gr layer, which in contrast to our findings was previously thought to be negligible and thus not important for the observed enhancement of interfacial DMI, is consistent with the energy spin-splitting we observe in the in-plane chiral spin component for Gr the $\uppi$ states in the thinner intercalated Co films.

\subsection {{\bf DFT calculations of Co-intercalated Gr/Ir interface}}

\begin{figure*}
\centering
\includegraphics[width=0.85\textwidth]{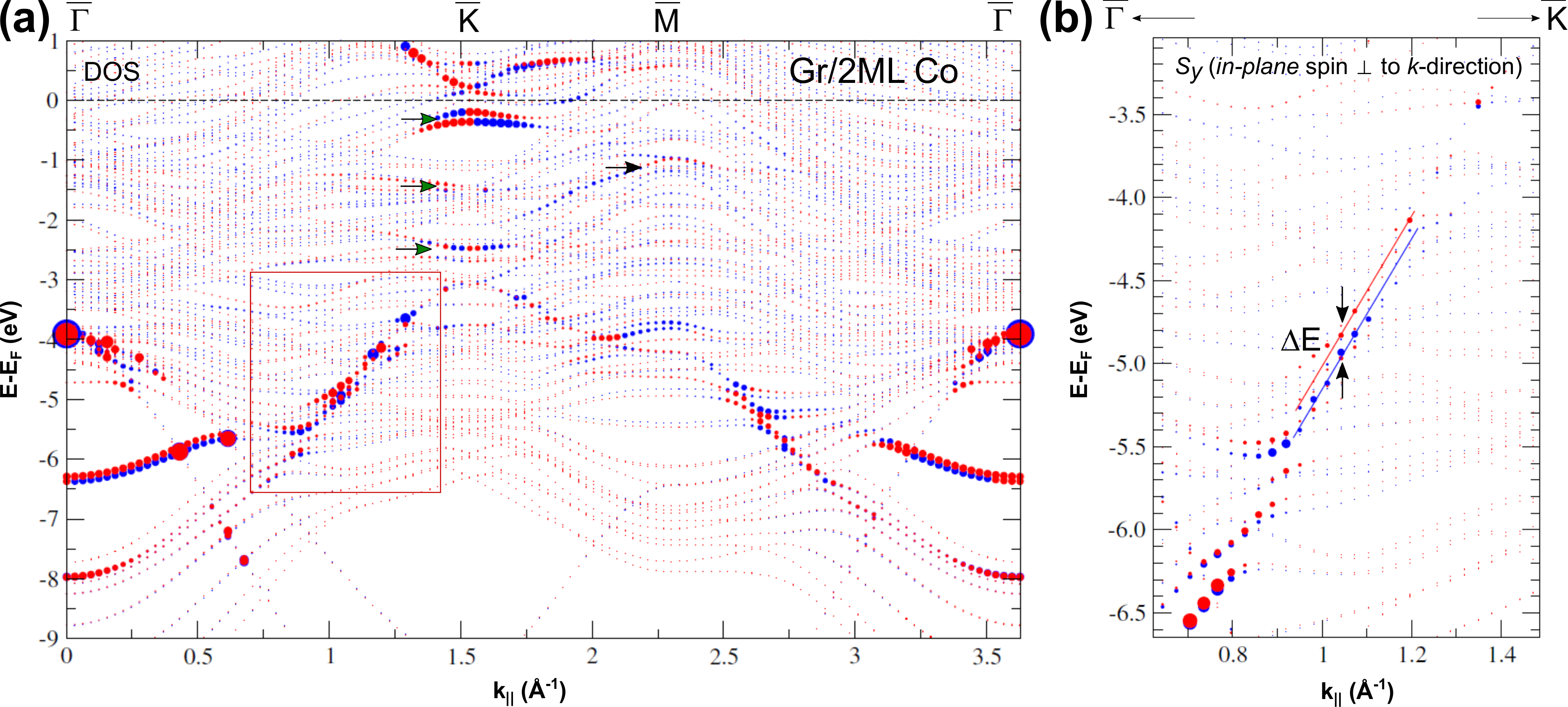}
\caption{{\bf Theoretical calculations of the electronic structure}. (a) Surface-projected band structure of Gr\,/2\,ML\,Co\,/Ir(111) as determined by DFT (see Methods) along the $\overline{\Gamma}$-$\overline{\mbox{K}}$-$\overline{\mbox{M}}$-$\overline{\Gamma}$ direction. The area of the circles is proportional to the product of the Gr and interface Ir DOS, while the color representation indicates the opposite orientations of the in-plane chiral spin component along the direction perpendicular to electron momentum. (b) Spin-resolved valence band structure corresponding to the calculated in-plane chiral spin orientations in the Gr layer. A close view of the energy-momentum dispersion of Gr $\uppi$ states along the $\overline{\Gamma}$-$\overline{\mbox{K}}$ direction is seen. The energy and momentum ranges correspond to the region of the Dirac cone with strong Gr character indicated by a rectangle in (a).}
\label{Fig7_DFT}
\end{figure*}

In an attempt to further verify our experimental findings, we performed DFT calculations for 2 ML of epitaxial Co intercalated in a Gr\,1${\times}$1\,/Ir(111) interface, with spin-orbit interaction fully taken into account. To reduce the complexity of the problem, in the calculations Gr\,1$\times$1 was expanded to match with the Ir lattice, and thus $\overline{\mbox{K}}$ and $\overline{\mbox{M}}$ high-symmetry points are referred to the Ir(111) SBZ. In this way, contributions from Gr moir\'e superlattice to the electronic structure are explicitly neglected. In Fig.~\ref{Fig7_DFT}(a)
we display the surface-projected band structure for Gr\,/2\,ML\,Co\,/Ir(111) as determined by DFT (see Methods) along the $\overline{\Gamma}$-$\overline{\mbox{K}}$-$\overline{\mbox{M}}$-$\overline{\Gamma}$ direction. Here, we highlight the contribution of the density of states (DOS) of Gr and Ir at the Gr/Co interface, while the color representation indicates the opposite orientations of the in-plane chiral spin component at the interface, which are perpendicular to electron momentum. Overall, the calculation shows qualitatively good agreement with the experimental results. The area of the circles is proportional to DOS\,(Gr)\,$\times$\,DOS\,(Ir), being thus indicative of the hybridization between Gr and Ir. More details on the spin- and surface-projected band structures are provided in SI, Fig.~S6. 

By directly comparing the DFT calculations with the experiment, it can be clearly seen that the Gr $\uppi$ bands are preserved and hybridized with Ir-Co states at the interface, resulting in a significant Gr-Ir interaction (energy-momentum region indicated by a rectangle in Fig.~\ref{Fig7_DFT}(a)). The main Dirac crossing point of the Gr $\uppi$ band at 
 the $\overline{\mbox{K}}$ point is located at an energy of E-E$_{\textrm{F}}$\,=\,-3 eV, in good agreement with the experimental results (see Fig.~\ref{Fig2_bands_comparison}(b) and supplementary Figs.~S1-S4). Similarly, in Fig.~\ref{Fig7_DFT}(a), clear signatures of the Gr-Ir hybridization can be also identified at -1.5 eV and in the vicinity of the E$_{\textrm{F}}$ around $\overline{\mbox{K}}$ (green arrows), where the strong C-Co hybridization gives rise to the characteristic mini Dirac cone as experimentally observed (Fig.~\ref{Fig4_FS_CE}(c)). Note that the calculations neglect the band shift due to electron doping, therefore the mini Dirac cone appears slightly above the E$_{\textrm{F}}$. Moreover, near the $\overline{\mbox{M}}$ point there is a small enhancement of the DOS (black arrow in Fig.~\ref{Fig7_DFT}(a)) consistent with the Ir-Co hybridization that is observed in the ARPES measurements (Figs.~\ref{Fig3_GM_hyb}, \ref{Fig4_FS_CE} and Fig.~S3 in SI).

In Fig.~\ref{Fig7_DFT}(b), we provide a closer view of the energy-momentum dispersion of Gr $\uppi$ states along the $\overline{\Gamma}$-$\overline{\mbox{K}}$ direction. Here, we show the calculated in-plane chiral spin component in the Gr layer within the energy and momentum ranges corresponding to the region indicated by a rectangle in Fig.~\ref{Fig7_DFT}(a). Noticeably, the chiral spin up and spin down orientations in Fig.~\ref{Fig7_DFT}(b) correspond those in Fig.~\ref{Fig7_DFT}(a). In both cases, there is a large energy spin-splitting in the in-plane chiral spin component between Gr $\uppi$ band states of opposite spin, confirming the crucial role of the spin-orbit interaction induced in the Gr layer for the enhancement of SOC at the interface. The spin-splitting is significantly pronounced in an energy region between E-E$_{\textrm{F}}$\,=\,-6 eV and -4 eV, that is, as the momentum approaches the $\overline{\mbox{K}}$ point, in very good agreement with our spin-resolved ARPES measurements (see Figs.~\ref{Fig5_SR_ARPES_fits} and \ref{Fig6_Bessy}). This spin-orbit splitting, which exceeds by far the one originating from SOC variations in the Co layer and is highlighted by red and blue solid lines in Fig.~\ref{Fig7_DFT}(b), is in quantitative agreement with the energy spin-splitting experimentally observed, that is, we recall, $\sim$\,100 meV. Consistent with this, a careful examination of the Gr $\uppi$ states in this energy region clearly shows that they are hybridized with the states localized at Ir (and Co) at the interface (larger DOS inside the rectangle in Fig.~\ref{Fig7_DFT}(a)). Thus, our theoretical calculations largely agree with the experimental results and demonstrate the strong electronic interaction between electronic states of Gr and those of the Ir-Co interface as well as a pronounced in-plane energy spin-splitting. In other words, the obtained results confirm the Rashba-like spin-texture induced by Ir in the Gr layer and shed light on the microscopic mechanism that is ultimately responsible for a significant enhancement of SOC at the Gr/Co interface in the experimental picture.

\section{CONCLUSION}

To summarize, through angle-resolved photoemission spectroscopy with and without spin resolution in combination with DFT calculations we have investigated the electronic origin of the large DMI at Gr/FM interfaces. To this end, we have performed a systematic experimental characterization of the electronic properties of high-quality Gr-based heterostructures containing atomically flat interfaces, as well as homogeneous epitaxial Co layers intercalated between Gr and a Ir(111) buffer layer grown on insulating substrates. Our findings reveal that, in contrast to conventional wisdom, spin-orbit effects induced in the Gr layer are of central importance to fundamentally understand the enhancement of SOC at Gr/FM interfaces. The effect is manifested by a large energy spin-splitting in the in-plane chiral component of spin for Gr $\uppi$ states, consistent with a Rashba spin texture and a spin-orbit splitting that exceeds by far the one originating from SOC variations in the Co layer. The experimental results are supported by DFT calculations pinpointing the tunable interaction between Gr and the Ir underlayer as the main source of SOC for C atoms. These findings are consistent with a large Rashba-SOC at the Gr/Co interface arising from the Gr layer in ultrathin intercalated Co films, and provide a convenient way to explore routes towards achieving unprecedented control of the interfacial properties. Our present findings taken altogether are highly relevant for the development of Gr-based memory and logic devices with advanced functionalities for future spintronic applications.

\section{METHODS}

{\bf Sample growth and characterization.} The samples were grown at the Molecular Beam Epitaxy (MBE) chamber at IMDEA Nanoscience (Madrid, Spain), and transferred to the Cassiop\'ee beamline end station at SOLEIL synchrotron (Gif-sur-Yvette, France) using an ultrahigh vacuum (UHV) suitcase at pressures below 5\,$\times$10$^{-10}$\,mbar. Gr-based epitaxial heterostructures were grown under UHV conditions on commercially available Al$_2$O$_3$(0001)-oriented oxide single crystals. The insulating substrates were \textit{ex-situ} annealed in air at 1370 K for 2 hours in order to obtain flat surfaces with large terraces prior to their insertion into the MBE chamber. Epitaxial (111)-oriented 30 nm thick Ir buffers were deposited by DC sputtering at 670 K in 8\,$\times$10$^{-3}$\,mbar Ar partial pressure with a deposition rate of 0.3 \AA/s. The epitaxial quality of the fabricated Ir layers was verified by LEED and XPS. Epitaxial Gr was subsequently grown at 1025 K by exposing the samples to ethylene gas at a partial pressure of 2\,$\times$10$^{-8}$\,mbar for 30 min. After cooling down to room temperature, Co was deposited by electron-beam evaporation on top of Gr/Ir/Al$_2$O$_3$(0001) with a deposition rate of 0.04 \AA/s, while its intercalation under Gr was promoted by a moderate thermal annealing. During this process, the sample was gradually heated up to 550 K while acquiring XPS spectra to verify in real time the complete intercalation of Co underneath the Gr layer. The XPS spectrum of Co was not modified by the presence of Co-C \cite{Ajejas2020} which rules out diffusion of C. The 4\,ML Co-intercalated sample for spin-resolved ARPES measurements was grown \textit{in-situ} at the spin-ARPES end station permanently installed at the U125-PGM beamline of BESSY-II. The Co source was carefully calibrated using a quartz crystal monitor.\\

{\bf Photoemission experiments.} ARPES measurements were carried out at room temperature at the Cassiop\'ee beamline HR-ARPES end station, and at the spin-ARPES station at the U125-PGM beamline of BESSY-II in Helmholtz-Zentrum Berlin. Photoemission spectra were acquired using linearly-polarized synchrotron light at photon energies of h$\upnu$\,=\,64~eV and 45 eV, respectively. The base pressure of the photoemission setups was better than 1\,$\times$10$^{-10}$\,mbar.  Emitted photoelectrons were detected with Scienta R4000 hemispherical analyzers up to acceptance angles of $\pm$15$^{\circ}$. The angular and energy resolutions were set to 0.1$^{\circ}$ and 5 meV, respectively.\\

{\bf Spin-resolved measurements.} Spin-resolved ARPES data in Fig.~\ref{Fig5_SR_ARPES_fits} were acquired at a photon energy of 64 eV at the spin-ARPES end-station of Cassiop\'ee beamline at SOLEIL synchrotron, using a SES2002 Scienta analyzer coupled to a Mini-Mott spin detector capable of detecting simultaneously both in-plane chiral (perpendicular to analyzer slit) and out-of-plane spin components (perpendicular to the surface plane). The energy resolution of the spin-resolved measurements was 230 meV, and the angular resolution was 3.6$^{\circ}$. A home-made Igor Wavemetrics Fitting procedure program was used to fit spectra in Fig.~\ref{Fig5_SR_ARPES_fits}. This procedure considers the Fermi level step and cutoff and its intrinsic broadening due to resolution and temperature, a Shirley background function and a multi-peak Voigt function to fit the several photoemision features. For each peak, the fitting parameters are related to their amplitude, width, energy and shape. The peak line-shape parameter can be modeled as a Voigt function to take into account the Gaussian broadening intrinsic to the experimental resolution. Spin-resolved data in Fig.~\ref{Fig6_Bessy} were acquired at the U125-PGM beamline of the BESSY-II  synchrotron light source, using a Scienta R4000 analyzer coupled to a Rice
University Mott-type spin polarimeter operated at 25 kV. Resolutions of the spin-resolved ARPES measurements were 45 meV (energy) and 0.75$^{\circ}$ (angular). The spin-resolved EDCs were fitted using Lorentzian peaks and a Shirley background, multiplied with a Fermi-function, and convoluted with a Gaussian broadening to account for the finite experimental resolution.\\

{\bf Theoretical calculations.} We used DFT in the generalized gradient approximation~\cite{PBE1996} employing the full-potential linearized augmented-plane-wave method as implemented in the {\sc Fleur} code~\cite{Kurz2004}. Relativistic effects, including spin-orbit coupling, were fully included. To calculate the electronic structure of Gr\,1${\times}$1\,/2\,ML\,Co\,/Ir(111), a seven layer Ir(111) film was used as substrate, and different stackings were considered as a starting point for the structures to relax. The lowest energy was obtained when one C atom was situated above the top Co and the other in a fcc position. To obtain the surface- and spin-projected band structures, spin-orbit coupling with an out-of-plane spin-quantization axis was considered, enabling non-collinear calculations to resolve the in-plane spin components in the different layers.

\section{ACKNOWLEDGMENTS}

This project has received funding from the FLAG-ERA JTC 2019 grant SOgraphMEM through the partner's national research agencies (Spain, AEI PCI2019-111867-2, France, ANR-19-GRF1-0001-07, Germany DFG-SOgraphMEM). IMDEA team  acknowledges support by the Community of Madrid (CM) through Project P2018/NMT-4321 (NANOMAGCOST), by MICIN/AEI through Projects RTI2018-097895-B-C42 (FUN-SOC), PGC2018-098613-B-C21 (SpOrQuMat), PID2021-122980OB-C52 (ECLIPSE-ECoSOx), EQC2019-006304-P, PID2020-116181RB-C31 (SOnanoBRAIN) and PID2021-123776NB-C21 (CONPHASE$^{\mathrm{TM}}$), and by the 'Severo Ochoa' Programme for Centres of Excellence in R\&D, MINECO grant CEX2020-001039-S. B.M.C., A.G. and I.A. acknowledge support from CM (PEJD-2019-PRE/IND-17048, PEJD-2017-PRE/IND-4690 and PEJD-2019-POST/IND-15343, respectively) and J.M.D. from MINECO (BES 2017-080617). We acknowledge SOLEIL for provision of synchrotron radiation facilities and we would like to thank CASSIOP\'EE beamline staff for assistance (experiment no. 20181593) and SEXTANTS beamline staff (in-house experiments). J.S.-B. acknowledges financial support from the Impuls- und Vernetzungsfonds der Helmholtz-Gemeinschaft under grant No. HRSF-0067. P.P. acknowledges fruitful discussions with Dr. Nicolas Jaouen, Dr. Jorge I. Cerd\'a (in memoriam) and Prof. Andr\'es Arnau.

%

\end{document}